\documentclass[preprints,article,accept,moreauthors,pdftex]{Definitions/mdpi} 

\usepackage{bm}
\usepackage{bbm}
\usepackage{listings}
\usepackage{subfigure}
\usepackage{amsmath}
\usepackage{amssymb}
\usepackage{booktabs}
\usepackage{hyperref}
\usepackage{fourier}
\usepackage[T1]{fontenc}
\usepackage[]{avant}
\usepackage[ruled, lined, linesnumbered, commentsnumbered, longend]{algorithm2e}

\hypersetup{colorlinks,citecolor=blue}

\firstpage{1} 
\pubvolume{xx}
\issuenum{1}
\articlenumber{1}
\pubyear{2020}
\copyrightyear{2020}
\history{}

\newcommand*\diff{\mathop{}\!\kern0pt\mathrm{d}}

\Title{Maximum Implied Variance Slope - Practical Aspects}
\Author{Fabien Le Floc'h}
\Author{Fabien {Le Floc'h} $^{1}$, Winfried Koller}

\AuthorNames{Fabien Le Floc'h}

\address{$^{1}$ fabien@2ipi.com}
\abstract{In the Black-Scholes model, the absence of arbitrages imposes necessary constraints on the slope of the implied variance in terms of log-moneyness, asymptotically for large log-moneyness. The constraints are used for example in the SVI implied volatility parameterization to ensure the resulting smile has no arbitrages. This note shows that those no-arbitrage contraints are very mild, and that arbitrage is almost always guaranteed in a large range of slopes where the contraints are enforced.}
\keyword{arbitrages; implied volatility; Black-Scholes; quantitative finance; pricing.}

\begin{document}
	
\section{Introduction}
Under the Black-Scholes model, a drift-less underlying asset price $F$ is a lognormal stochastic process :
\begin{equation*}
	\diff F = \sigma F \diff W\,,
\end{equation*}
where $W$ is a Brownian motion and $\sigma$ is the Black-Scholes volatility. 
The price of a European call option contract of maturity $T$ and strike $K$ is known in closed form and reads \citep{black1976pricing}
\begin{align}
	C(K,\sigma) &= B(0,T) \left[ F(0,T)\Phi\left(d_1\right) - K\Phi\left(d_2\right)\right] \,,
\end{align}
where $\Phi$ is the cumulative normal distribution function, $B(0,T)$ is the discount factor to maturity and 
\begin{align*}
	d_1 = \frac{1}{\sigma\sqrt{T}}\ln\frac{F(0,T)}{K} + \frac{1}{2}\sigma \sqrt{T}&\,,\quad d_2 = d_1 - \sigma\sqrt{T}\,.
\end{align*}

In practice, the underlying asset is a forward rate, such as a forward swap rate for an interest rate swaption, or a forward foreign exchange rate, or a stock forward price. 
The market prices of European options imply different values of the Black-Scholes volatility, depending on the strike and the time to maturity of the option considered.

Roger Lee proves that the Black-Scholes implied volatility asymptotics can not be arbitrary in \cite{lee2004moment}: the implied distribution does not have finite moments if the square of the implied volatility $v(y,T)=\sigma^2(y,T)$ grows faster than $2|y|/T$ where $y$ is the log-moneyness $y=\ln\frac{K}{F}$, $K$ is the option strike, $F$ the underlying asset forward to maturity $T$ and $\sigma$ is the Black-Scholes implied volatility for the given moneyness and maturity.
As a consequence, any implied variance extrapolation must be at most linear in the wings, a fact of great practical interest.
Earlier, \citet{hodges1996arbitrage} derived simpler bounds on the Black-Scholes implied volatility based on the option slopes in  and \citet{gatheral2000rational} suggested that for many economically reasonable stochastic volatility plus jump models, the Black-Scholes implied variance is asymptotically linear.

In this note, we show that the implied volatility may contain arbitrages at very large strikes even though the Lee moments bounds are enforced. We start by presenting a concrete example based on the SVI parameterization of \citet{gatheral2004parsimonious} where call option prices increase significantly on a large range of strikes.  We then propose a practical criteria which indicates that the bounds on the implied variance slope must be much smaller.

Increasing call option prices may be particularly problematic for pricing contracts based on the replication of \citet{neuberger1994log,carr2001towards}, such as variance swaps, as it may significantly distort the obtained price.

\section{Arbitrages in SVI at very large strikes}

Inspired by the behavior of many stochastic volatility models with jumps at large strikes, \citet{gatheral2004parsimonious} proposed the SVI parameterization of the implied variance, widely adopted by pratitioners despite known shortcomings:
\begin{equation}
	v(y,T) = a  + b \left[\rho (y-m) + \sqrt{(y-m)^2 + s^2}\right]\,,
\end{equation}
where $(a,m) \in \mathbb{R}^2$, $b \geq 0$, $s > 0$, $\rho \in [-1,1]$ are the model parameters used to fit the market implied volatilities.

In order to prevent call spread and butterfly spread arbitrages, \citet{gatheral2004parsimonious} recommends to enforce the condition
\begin{equation}
	b(1+|\rho|) < \frac{4}{T}\,.\label{gatheral_slope}
\end{equation}
This is not enough to obey the asymptotics of \citet{lee2004moment}.
Indeed, the left and right asymptotes read $v_L = a - b (1-\rho)(k-m)$ and $v_R = a + b (1+\rho)(k-m)$. And thus the Lee moment formula implies
\begin{equation}
	b(1+|\rho|) < \frac{2}{T}\,.\label{lee_slope}
\end{equation}
Furthermore, \citet{hodges1996arbitrage} shows that $d_1 \to -\infty$ for $k \to +\infty$, when the inequality is strict. This means that the call option price is guaranteed to go to zero for large strikes. When the implied variance has an asymptotic slope of $2$ for large log-moneyness, there are several cases to consider but the call prices may not converge to zero anymore. When the asymptotic slope is strictly larger that two, we have $d_1 \to +\infty$ and arbitrage is guaranteed.

\begin{figure}[h]
	\centering{
		\subfigure[\label{fig:svi_vols_slopes}Implied volatility.]{
			\includegraphics[width=.49\textwidth]{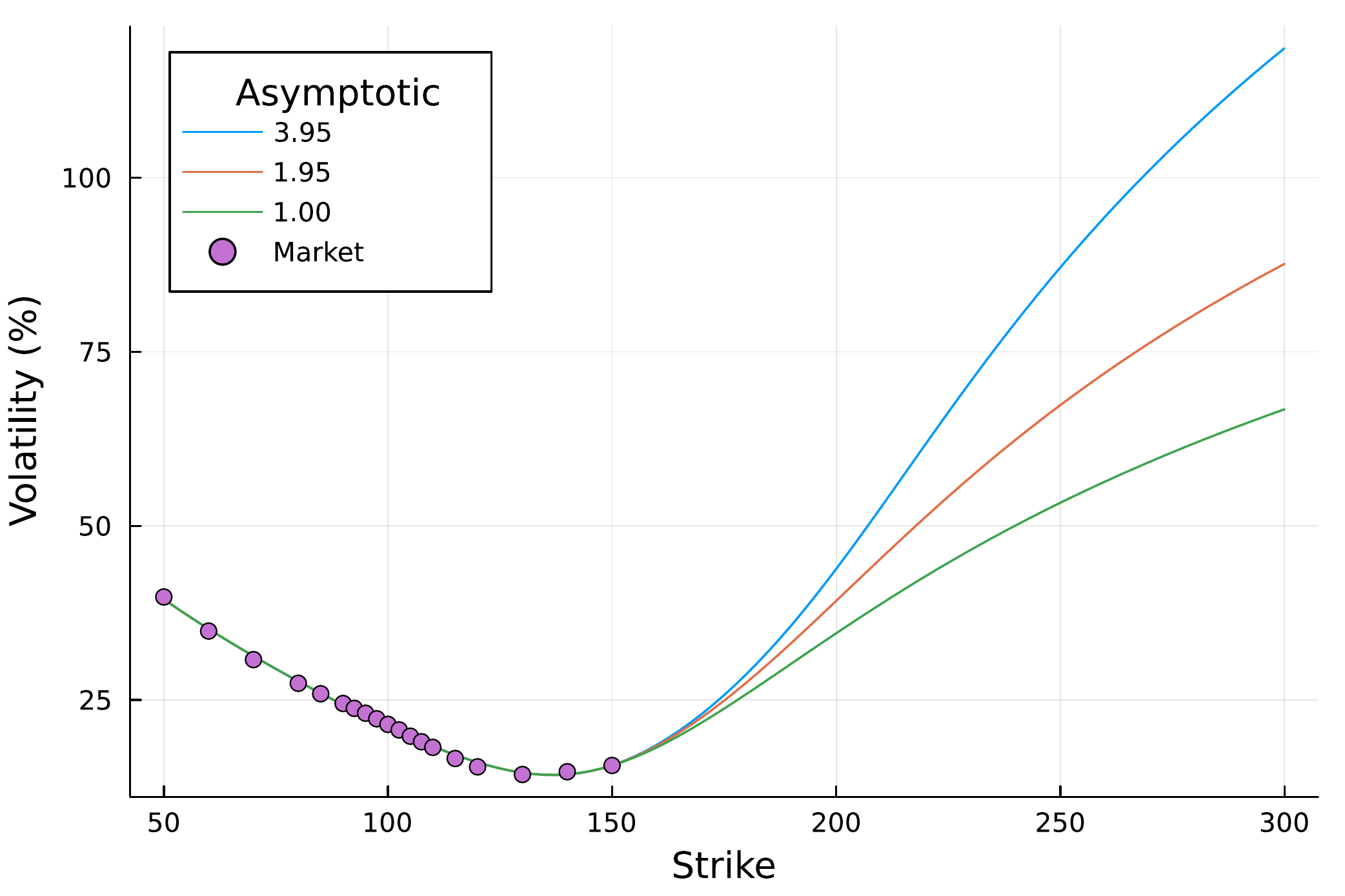}}
		\subfigure[\label{fig:svi_prices_slopes}Call price]{
			\includegraphics[width=.49\textwidth]{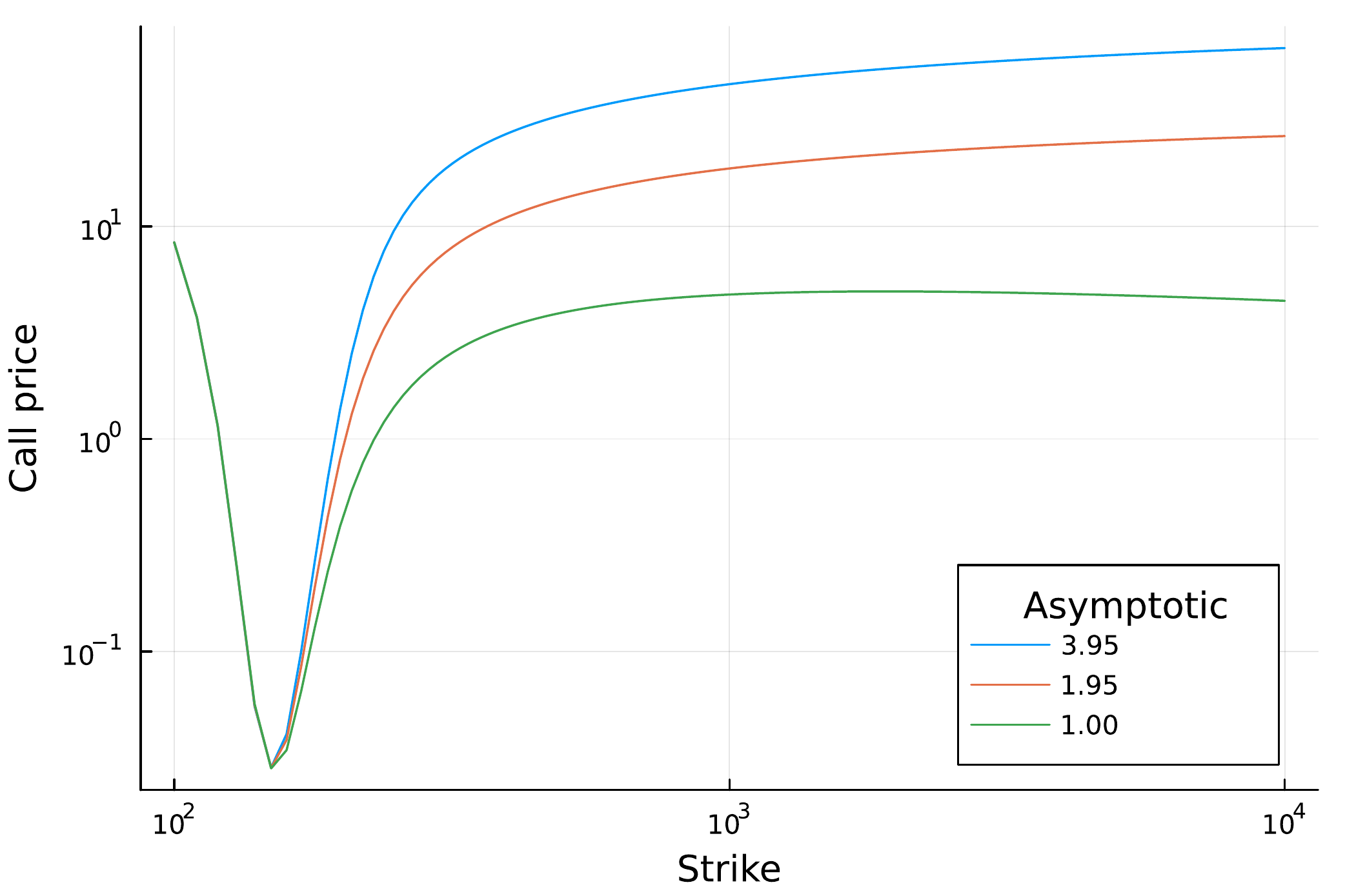} }
		\caption{Implied volatility smile of maturity 1 year, and call price for SVI parameterizations with different asymptotic slopes. Call prices are still increasing at very large strikes.}}
\end{figure}

The above are necessary conditions. It is thus of no surprise to find examples where the implied volatilities obey the bounds and yet have call-spread or butterfly-spread arbitrages. It is however more surprising to find examples where, at very large strikes arbitrages still occur.
Such examples are given in Table \ref{tbl:svi_params_slope} and Figures \ref{fig:svi_vols_slopes}, \ref{fig:svi_prices_slopes}.

\begin{table}[h]
	\caption{SVI parameters corresponding to calibrations to the options of Table \ref{tbl:bad_vol} using different constraints on the asymptotic slope.\label{tbl:svi_params_slope}}
	\centering{
	\begin{tabular}{cccccr}\toprule
		$a$ & $b$ & $s$ & $\rho$ & $m$ & Asymptotic slope\\\midrule
		-0.152555 & 2.073631 & 0.195700 & 0.904871 & 0.729450 & 3.95 \\
		-0.136299 & 1.072730 & 0.253555 & 0.817793  &0.673280  & 1.95 \\
	    -0.112306 & 0.596259 & 0.302274 & 0.677123 & 0.590297 &1.00\\ \bottomrule
	\end{tabular}}
\end{table}

In the case where $b(1+|\rho|)=1.95$, the call option prices are still increasing for $K > 10^6 F(0,T)$.
This stays true when the asymptotic slope $S$ is lower yet. For $S=1$, the price starts decreasing only around $K=1900$, quite far off the forward price $F=100$, for a maturity $T=1$ year, while, in the absence of arbitrage, the call prices must be monotonically decreasing.

\section{What would be an acceptable limit on the asymptotic slope?}
For very large strikes, we would like the call option price to be realistic, typically close to zero. We may consider strike much larger than what the market will ever use, $K_{\max}=10^6 F$ and impose that the price must be below $C_{\max} = 10^{-4}F$ as a very conservative first estimate. 
Neglecting\footnote{The contribution of $d_2$ could easily be included, we may use an implied volatility solver to find the limiting slope $S$ then.} the contribution due to $d_2$, this rule leads to an asymptotic slope $S$ verifying \begin{equation}C_{\max} \approx F(0,T) B(0,T) \Phi\left(\frac{-y_{\max}+\frac{S y_{\max}T}{2}}{\sqrt{S y_{\max}T}}\right)\,,\end{equation}
with $y_{\max} = \ln K_{\max} / F(0,T)$. This leads to a quadratic equation: 
\begin{equation}
	\frac{1}{2}\sqrt{ST}^2 - \frac{1}{\sqrt{y_{\max}}}\Phi^{-1}\left(\frac{C_{\max}}{B(0,T)F(0,T)}\right) \sqrt{ST} - 1 \approx 0\,.
\end{equation}
In our case, we end up with $S \approx 0.5355$. In particular this is much lower than the limit imposed by Lee moments formula $S=2$, independently of the implied volatility parameterization. Conversely, arbitrage is very likely if a parameterization leads to $S > \frac{0.5355}{T}$.

More realistically, we would want the call option prices to be not greater than the market call option price with largest quoted strike. In our example it is $K=150$ with volatility 15.6\%. This leads to $C_{\max}=0.02793$. We may also consider a lower maximum strike, for example $K_{\max} = 1500$, and this would lead to $S \approx 0.1869$.

\begin{figure}[H]
	\centering
	\includegraphics[width=0.8\textwidth]{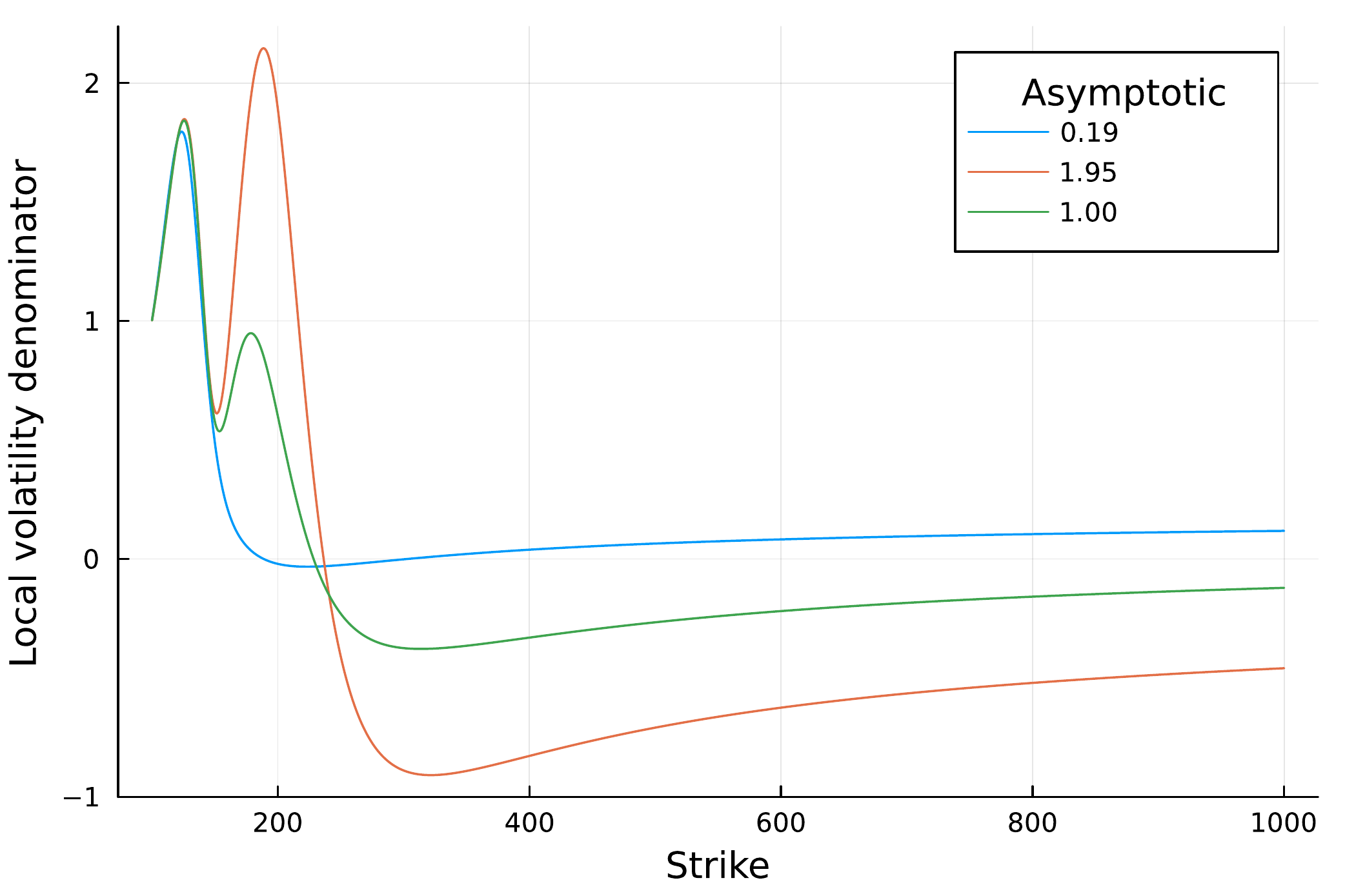}
	\caption{Local volatility denominator for SVI calibrated to the market data of Table \ref{tbl:bad_vol} using various asymptotic slope constraints $S$. With $S=0.19$, the density stays positive beyond strikes $K > 320$ but arbitrages are present within $(190,320)$.}
	\label{fig:vol_svi_dens_bad}
\end{figure}

On this example, the practical SVI slope limit is thus more than 10 times smaller than what the Lee moment formula suggests. Even with such a low slope, there are still arbitrages in the SVI representation, but those are located not too far in the wings. In Figure \ref{fig:vol_svi_dens_bad}, butterfly spread arbitrages correspond to zones where the (Gatheral) local volatility denominator is negative. The local volatility denominator $g$ in terms of the total implied variance $w(y,T)=v(y,T)T$ reads \citep{gatheral2014arbitrage}
\begin{equation}
g(y)=	1-\frac{y}{w}\frac{\partial w}{\partial y}+\frac{1}{4}\left(\frac{\partial w}{\partial y}\right)^2 \left(-\frac{1}{4} - \frac{1}{w} + \frac{y^2}{w^2}\right) + \frac{1}{2}\frac{\partial^2 w}{\partial y^2}\,.
\end{equation}

\section{Conclusion}
The asymptotic bounds on the slope of the implied variance for large strikes impose important limitations on the parameterizations of the implied volatility. Actual practical bounds are however typically much lower than what is suggested by the theoretical formulae.

\externalbibliography{yes}
\bibliography{maximum_slope.bib}
\appendixtitles{no}
\appendix
\section{Example market data}
\begin{table}[h]
	\caption{Market implied volatilities in ACT/365 for an option of with 365 days to maturity $T$ on an asset of forward $F(0,T)=1.0$.\label{tbl:bad_vol}}
	\centering{	{
			\begin{tabular}{lcccccccccc}\toprule
				Strike&		  0.5 & 0.6&
				0.7&
				0.8&
				0.85&
				0.9&
				0.925&
				95&
				0.975&
				1.0
				\\
				Volatility (\%)&		 39.8&
				34.9&
				30.8&
				27.4&
				25.9&
				24.5&
				23.8&
				23.1&
				22.3&
				21.5
				\\\midrule
				Strike & 	1.025&
				1.05&
				1.075&
				1.1&
				1.15&
				1.2&
				1.3&
				1.4&
				1.5&\\
				Volatility (\%) & 20.7&
				19.8&
				19.0&
				18.2&
				16.6&
				15.4&
				14.3&
				14.7&
				15.6& \\
				\bottomrule
	\end{tabular}}}
\end{table} 

\end{document}